\documentclass[sigconf]{acmart}
\usepackage{multirow}
\usepackage{array}
\usepackage{enumitem}
\AtBeginDocument{%
  }

\setcopyright{acmlicensed}

\copyrightyear{2026}
\acmYear{2026}
\setcopyright{cc}
\setcctype{by}
\acmConference[KDD '26]{Proceedings of the 32nd ACM SIGKDD Conference on Knowledge Discovery and Data Mining V.1}{August 09--13, 2026}{Jeju Island, Republic of Korea}
\acmBooktitle{Proceedings of the 32nd ACM SIGKDD Conference on Knowledge Discovery and Data Mining V.1 (KDD '26), August 09--13, 2026, Jeju Island, Republic of Korea}
\acmPrice{}
\begin{document}

\title{InfoDCL: Informative Noise Enhanced Diffusion Based Contrastive Learning}

\author{Xufeng Liang}
\email{xfliang@bit.edu.cn}
\affiliation{%
  \institution{Beijing Institute of Technology}
  \city{Beijing}
  \country{China}
}

\author{Zhida Qin}
\email{zanderqin@bit.edu.cn}
\affiliation{%
  \institution{Beijing Institute of Technology}
  \city{Beijing}
  \country{China}
}

\author{Chong Zhang}
\email{zhangchong@xjtu.edu.cn}
\affiliation{%
  \institution{Xi'an Jiaotong University}
  \city{Xi'an}
  \country{China}
}

\author{Tianyu Huang}
\email{huangtianyu@bit.edu.cn}
\affiliation{%
  \institution{Beijing Institute of Technology}
  \city{Beijing}
  \country{China}
}

\author{Gangyi Ding}
\email{dgy@bit.edu.cn}
\affiliation{%
  \institution{Beijing Institute of Technology}
  \city{Beijing}
  \country{China}
}

\renewcommand{\shortauthors}{Xufeng Liang et al.}

\begin{abstract}
Contrastive learning has demonstrated promising potential in recommender systems. Existing methods typically construct sparser views by randomly perturbing the original interaction graph, as they have no idea about the authentic user preferences. Owing to the sparse nature of recommendation data, this paradigm can only capture insufficient semantic information. To address the issue, we propose InfoDCL, a novel diffusion-based contrastive learning framework for recommendation. Rather than injecting randomly sampled Gaussian noise, we employ a single‐step diffusion process that integrates noise with auxiliary semantic information to generate signals and feed them to the standard diffusion process to generate authentic user preferences as contrastive views. Besides, based on a comprehensive analysis of the mutual influence between generation and preference learning in InfoDCL, we build a collaborative training objective strategy to transform the interference between them into mutual collaboration. Additionally, we employ multiple GCN layers only during inference stage to incorporate higher-order co-occurrence information while maintaining training efficiency. Extensive experiments on five real-world datasets demonstrate that InfoDCL significantly outperforms state-of-the-art methods. Our InfoDCL offers an effective solution for enhancing recommendation performance and suggests a novel paradigm for applying diffusion method in contrastive learning frameworks.
\end{abstract}

\begin{CCSXML}
<ccs2012>
 <concept>
  <concept_id>00000000.0000000.0000000</concept_id>
  <concept_desc>Do Not Use This Code, Generate the Correct Terms for Your Paper</concept_desc>
  <concept_significance>500</concept_significance>
 </concept>
 <concept>
  <concept_id>00000000.00000000.00000000</concept_id>
  <concept_desc>Do Not Use This Code, Generate the Correct Terms for Your Paper</concept_desc>
  <concept_significance>300</concept_significance>
 </concept>
 <concept>
  <concept_id>00000000.00000000.00000000</concept_id>
  <concept_desc>Do Not Use This Code, Generate the Correct Terms for Your Paper</concept_desc>
  <concept_significance>100</concept_significance>
 </concept>
 <concept>
  <concept_id>00000000.00000000.00000000</concept_id>
  <concept_desc>Do Not Use This Code, Generate the Correct Terms for Your Paper</concept_desc>
  <concept_significance>100</concept_significance>
 </concept>
</ccs2012>
\end{CCSXML}

\ccsdesc[500]{Information systems~Recommender systems}

\keywords{Collaborative Filtering, Diffusion Model, Contrastive Learning}

\maketitle

\section{Introduction}
With the rapid expansion of digital information, recommender systems have become essential tools for filtering as well as distributing vast amounts of data. As user interaction records continue to accumulate at an unprecedented rate, there is a growing need for recommendation models that can accurately and efficiently uncover users' underlying preferences. In recent years, contrastive learning has gained significant attention in the research community due to its powerful ability to learn discriminative representations by contrasting positive and negative samples. Compared to traditional supervised learning methods, contrastive learning can leverage abundant unlabeled data to capture fine-grained similarities, making it particularly well-suited for recommendation tasks where explicit feedback is often sparse and noisy.

Contrastive learning has emerged as a powerful self‑supervised paradigm in recommendation\cite{hgclcf,hgcf,lightgnn,cluia,intent,uniform,dual,robust,sub,cgi,xsimgcl,disenpoi,mlpcl,ragcl,clbm}, enabling robust user and item representations. As one of the pioneering works, SGL\cite{sgl} incorporates it into graph‑based collaborative filtering by constructing multiple graph augmentations to enhance representation discrimination. NCL\cite{ncl} expands on this by forming contrastive pairs with both structural and semantic neighbors to better utilize neighborhood information. Besides, SimGCL\cite{simgcl} proposes a model‑agnostic framework that injects noise into embeddings, facilitating robust learning without handcrafted augmentations. Building on these, CoGCL\cite{cogcl} leverages discrete collaborative codes and virtual neighbors to create semantically meaningful contrastive views and strengthen collaborative signals. In the field of generative models, DiffMM\cite{DiffMM} attempts to integrate a modality-aware graph diffusion model with cross-modal contrastive learning to align multimodal item contexts with collaborative relations, generating modality-aware user-item graphs via diffusion process and enhancing representations. Numerous other models\cite{gfdiff,mvideorec,ownership,cmdl,eager,joint,separated,hierarchical} also explore diverse augmentation strategies and contrastive objectives, further advancing the contrastive recommendation landscape.

\textbf{Motivations.} Though achieving impressive progress, current contrastive learning recommendation models still suffer from a core issue - they have no idea about the distribution of authentic user preferences - so they just focus on making perturbations on existing interaction relationships to generate sparser views. And other improved methods, such as SimGCL\cite{simgcl}, point out that it is unnecessary to perturb the interaction graph structure; instead, simply adding noise to the embeddings can achieve the effect of distribution uniformity. One key drawback of these approaches is that, owing to the sparsity of recommendation data, the views constructed solely from existing information can only reflect very limited user preferences. Therefore the user preferences captured by model through the contrastive-learning paradigm are consequently insufficient. Although some methods (e.g., DiffMM\cite{DiffMM}) introduce auxiliary data to generate views, they mainly align views derived from heterogeneous sources. However, due to substantial distributional gaps and potential noise within these auxiliary modalities, such alignments are often insufficient for learning accurate user preferences. 

Different from the methods that merely add perturbation or inject noise on interaction data, what we desire is to generate views that authentically reflect user preferences. Owing to the impressive ability to model complex data distributions and generate highly realistic outputs of generative model, we propose to utilize the powerful generative capacity of the diffusion paradigm to create preference‑aware views.

\textbf{Challenges.} The key to alleviating this problem lies in leveraging the diffusion paradigm to generate contrastive views that capture the semantics in auxiliary information while preserving the interaction relationships in embeddings and integrating them into the same semantic space. Nevertheless, to achieve these goals encounter two challenges: (i) \textit{How to excavate desired user preferences while avoiding the destruction of the co-occurrence relationships in inherent sparse data?} Owing to the sparsity of recommendation interactions, the straightforward transfer of diffusion paradigm from image synthesis, wherein the original interaction or item representations undergo randomly sampled Gaussian noise corruption, fails to generate the semantically rich embeddings required. (ii) \textit{How to collaboratively optimize the generation and preference learning?} According to our empirical observation and experiments, the training objectives between diffusion process and preference learning can interfere with each other, resulting in insufficient optimization.

\textbf{Contributions.} In this paper, we propose a novel and flexible \textit{\underline{info}rmative noise enhanced \underline{d}iffusion \underline{c}ontrastive \underline{l}earning framework}, which is called \textbf{InfoDCL}. Our framework is capable of constructing an independent contrastive learning channel for each type of semantic auxiliary information, and jointly training them in a coordinated strategy to achieve optimal performance. For each type of auxiliary information, we first design a \textbf{PSNet} to simulate the single-step diffusion process, which effectively generates informative noise enriched with semantics. The resulting informative noise is then injected into the diffusion forward process and subsequently denoised to generate the desired user preferences. For the embeddings generated from each channel, we contrast them with the shared, initialized item representations to better align the distribution of the item representations with authentic user preferences. We also build a \textbf{collaborative training objective strategy} that optimizes reconstruction loss, contrastive loss and BPR loss simultaneously to enhance learning effect. Additionally, we only utilize multiple GCN layers in the inference phase to further incorporate higher-order co-occurrence information, which eliminates the need for convolution operations during training, thereby significantly improving efficiency. We conduct extensive experiments on five real-world datasets and verify the superiority of our InfoDCL framework. Our contributions can be summarized as follows.
\begin{itemize}[leftmargin=*]
    \item We propose a novel and flexible contrastive learning framework that substantially modifies the construction of contrastive views to address the sparsity of recommendation data, and can be extended to incorporate various types of auxiliary information.
    \item We design a diffusion-based process that efficiently injects auxiliary semantic information into Gaussian noise, producing informative noise that better enhances embeddings generation to reflect authentic user preferences.
    \item We build a collaborative training objective strategy which transforms the interference between generation and preference learning into mutual collaboration, thereby substantially improving the model's learning ability and adaptability.
    \item We conduct evaluations on five real-world interaction datasets. Results show that our model significantly outperforms other baseline methods. Apart from this, we also perform empirical studies and offer theorems to improve the interpretability of our framework.
\end{itemize}

\section{Problem Definition}
\noindent $\bullet$ \textbf{Collaborative Graph.} In a recommender system, the input can be modeled as a binary interaction graph $\mathcal{G} = (\mathcal{U} \cup \mathcal{I}, \mathcal{E})$, where $\mathcal{U} = \{\boldsymbol{u}_1, \boldsymbol{u}_2, ..., \boldsymbol{u}_M\}$ denotes the set of users, and $\mathcal{I} = \{\boldsymbol{i}_1, \boldsymbol{i}_2, ..., \boldsymbol{i}_N\}$ denotes the set of items. The edge set $\mathcal{E}$ represents observed interactions, where an edge $(\boldsymbol{u}_m, \boldsymbol{i}_n) \in \mathcal{E}$ implies that user $\boldsymbol{u}_m$ has interacted with item $\boldsymbol{i}_n$. These user-item interactions can be captured using an adjacency matrix $\mathbf{A} \in \mathbb{R}^{M \times N}$, with $M$ and $N$ being the total numbers of users and items respectively. For any entry $\mathbf{A}_{mn}$, its value is $\boldsymbol{1}$ if there is an interaction between user $\boldsymbol{u}_m$ and item $\boldsymbol{i}_n$, and $\boldsymbol{0}$ otherwise. Furthermore, to incorporate rich semantic information to guide the generation process, we introduce the auxiliary metadata $\mathbf{m}$.

\noindent $\bullet$ \textbf{Task Formulation.} Given this graph, our task is to learn a function $\boldsymbol{f}$ that can estimate the likelihood of potential future interactions between users and items. For every user $\boldsymbol{u}_m$, we seek to produce a personalized ranking over the set of items $\{\boldsymbol{i}_n| (\boldsymbol{u}_m, \boldsymbol{i}_n) \notin \mathcal{E}\}$ that the user has not interacted with, according to predicted scores. The function $\boldsymbol{f}$ operates on an interaction graph augmented with auxiliary metadata $\mathcal{G}^{\mathcal{A}} = (\mathcal{G}, \boldsymbol{\mathbf{m}})$, and the prediction is formulated as $\hat{\boldsymbol{y}}_{u} = f(\mathcal{G}^{\mathcal{A}})$.

\section{Methodology}
\begin{figure*}[t!]
    \centering
    \captionsetup{skip=0pt}
    \includegraphics[width=\linewidth]{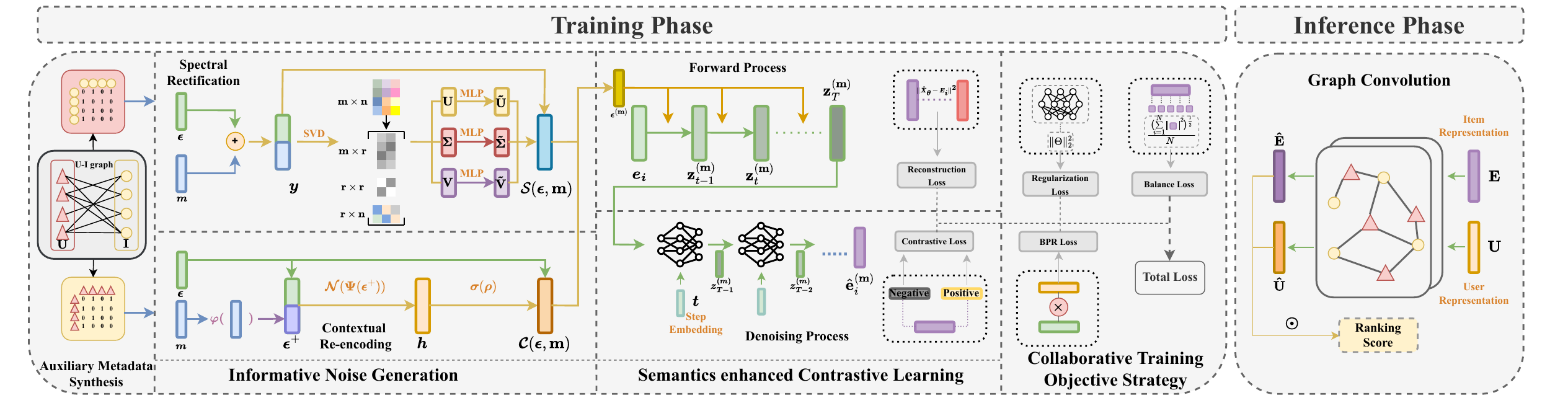}
    \caption{The overall architecture of our proposed InfoDCL, which involves informative noise generation by applying SVD to approximate a single-step diffusion process, where randomly sampled Gaussian noise and auxiliary metadata with semantic information are fused to produce informative noise. Then this signal is utilized in the standard diffusion paradigm to generate item embeddings with authentic user preferences. The collaborative training objective strategy continuously optimizes generation, preference learning and contrastive learning tasks. During inference, the GCN is incorporated to enhance representations with higher-order co-occurrence information, improving training efficiency.}
    \label{fig:oa}
\vspace{-0.1in}
\end{figure*}
In this section, we introduce InfoDCL, a framework that reconstructs randomly sampled Gaussian noise into semantically meaningful signals, which are injected in the diffusion forward process to produce embeddings with richer user preferences. Our framework is flexible and extensible, allowing it to incorporate various types of auxiliary information to jointly enhance the learning of diverse user preferences.

For each type of auxiliary information, in order to accurately and efficiently extract the semantic information it contains, based on the generative idea of the diffusion paradigm, we first utilize single-step diffusion to generate informative noise. This signal is used in the subsequent diffusion process to replace Gaussian noise, producing item embeddings that better capture user preferences. Our intuition is to train the diffusion process to excavate the features of items that users are interested in, thereby constructing views with richer semantics. For optimization, we adopt a collaborative training strategy where each type of auxiliary information's reconstruction and contrastive losses are jointly optimized with the BPR loss. Notably, to further capture higher-order co-occurrence patterns while improving training efficiency, we innovatively incorporate GCN exclusively at the inference stage. The following subsections detail all components.

\subsection{Informative Noise Generation}
Owing to the high sparsity inherent in interaction, randomly sampled Gaussian noise tends to corrupt the already limited latent user preference signals. To generate more informative noise for the diffusion process, we design a module named \textbf{Preference Signal Network (PSNet)}, which integrates auxiliary signals and leverages the decomposition ability of singular value decomposition (SVD) to produce informative noise. Motivated by the Davis–Kahan theorem\cite{stewart1990stochastic,xie2023overlooked}, this process indeed simulates a single-step diffusion.

Specifically, PSNet is built upon two core modules. The first, spectral rectification, approximates a single-step diffusion perturbation and reconstruction process by applying SVD to the input signal, which is composed of Gaussian noise and auxiliary information. The second, contextual re-encoding, enriches the semantic expressiveness by fusing it with auxiliary metadata to mitigate the incomplete approximation. Together, the two modules produce the informative noise with enhanced semantic content.
\subsubsection{\textbf{Auxiliary Metadata Synthesis}}
\
\newline
\label{sec:metadata generation}
Since the randomly sampled noise itself lacks semantic information, we respectively construct a user similarity graph and an item similarity graph to create auxiliary metadata that co-participates in the generation of informative noise, thereby injecting user-preference information. Concretely, we rely solely on the user–item interactions contained in the training set to construct a user social-relation graph and an item feature-similarity graph, based on which we aggregate the LightGCN pre-trained embeddings to obtain the auxiliary metadata.

Noticeably, the auxiliary metadata is extensible: when a dataset provides authentic social relations, knowledge graph links, or multimodal attributes, these can likewise be incorporated to enrich the semantic content of metadata.
\subsubsection{\textbf{Spectral Rectification}}
\ 
\newline
The first key component of PSNet is \textbf{spectral rectification}, which aims to approximate the single-step diffusion generation process by performing singular value decomposition on the input signal and encoding the obtained singular vectors to learn the semantics in the auxiliary metadata. Previous studies\cite{stewart1990stochastic,xie2023overlooked,usefulVariantDK} have confirmed the effectiveness of this process. In practice, before we compute the SVD process, the input signal $\mathbf{y}$ is acquired by a combination of noise $\boldsymbol{\epsilon}$ and metadata $\mathbf{m}$. The total process can be formulated as follows:
\begin{equation}
\mathbf{y} = \boldsymbol{\epsilon} + \mathbf{m}, \quad \mathbf{y} = \mathbf{U} \boldsymbol{\Sigma} \mathbf{V}^{\!\top}, \qquad d \ll D
\end{equation}
where $\mathbf{U} \in \mathbb{R}^{1 \times d}$, $\mathbf{V} \in \mathbb{R}^{D \times d}$ and $\boldsymbol{\Sigma} \in \mathbb{R}^{d \times d}$ is the singular vectors. These three are subsequently fed into distinct MLP layers, thereby emulating the feature-generation process of the diffusion paradigm:
\begin{equation}
\tilde{\mathbf{U}} = \mathrm{MLP}(\mathbf{U}),\quad
\tilde{\mathbf{V}} = \mathrm{MLP}(\mathbf{V}^{\!\top}_{1:}),\quad
\tilde{\boldsymbol{\Sigma}} = \mathrm{MLP}(\operatorname{diag}(\boldsymbol{\Sigma})),
\end{equation}
where $\tilde{\mathbf{U}}, \tilde{\mathbf{V}}, \tilde{\boldsymbol{\Sigma}} \in \mathbb{R}^{d}$ and $\operatorname{diag}$ extracts the singular values from the diagonal of the matrix. Since each of the three singular values is transformed through a distinct MLP, we need to bring them back into the same semantic space. Therefore, we then concatenate them to recover the signal:
\begin{equation}
\mathbf{g} = \Phi([\tilde{\mathbf{U}}\,\|\,\tilde{\mathbf{V}}\,\|\,\tilde{\boldsymbol{\Sigma}}]) \in \mathbb{R}^{D},
\end{equation}
where $\Phi : \mathbb{R}^{3d} \to \mathbb{R}^{D}$ denotes a non-linear mapping. Finally, we apply a residual connection and get the output of spectral rectification:
\begin{equation}
\mathcal{S}(\boldsymbol{\epsilon}, \mathbf{m}) = \mathbf{y} + \operatorname{diag}(\tanh(\boldsymbol{\alpha})) \cdot \mathbf{g},
\end{equation}
where $\boldsymbol{\alpha} \in \mathbb{R}^{D}$ is a learnable parameter vector to control the residual scale. Owing to the observation that the singular vectors obtained through SVD of the input and output signals in a single-step diffusion tend to be similar in structure but may be in opposite directions, the current output $\mathcal{S}(\boldsymbol{\epsilon}, \mathbf{m})$, although closely approximating the diffusion-generated result, is still constrained by the limitations of SVD-based simulation. This necessitates further enhancement of the spectral rectification process.

\subsubsection{\textbf{Contextual Re-encoding}}
\ 
\newline
To further fuse noise and auxiliary metadata, we propose a \textbf{contextual re-encoding} module to bridge the discrepancy between the Gaussian noise and the informative noise. This module aims to broaden the width of PSNet to learn the residual between the output $\mathcal{S}(\boldsymbol{\epsilon}, \mathbf{m})$ and ideal informative noise. Specifically, we form an augmented input by combining them:
\begin{equation}
\boldsymbol{\mathbf{\epsilon}}^{+} = \boldsymbol{\epsilon} + \boldsymbol{\varphi}(\mathbf{m}) \in \mathbb{R}^{\mathbf{D}},
\end{equation}
After that, a linear transformation is employed to further integrate the semantic information and noise contained in the metadata:
\begin{equation}
\mathbf{h} = \mathcal{N}(\Psi(\boldsymbol{\epsilon}^{+})) \in \mathbb{R}^{\mathbf{D}},
\end{equation}
where $\Psi$ is a linear function and $\mathcal{N}$ is layer normalization. Meanwhile, a residual scaling factor $\boldsymbol{\rho}$ regulates the residual strength. Finally, the contextual re-encoding output is:
\begin{equation}
\mathcal{C}(\boldsymbol{\epsilon}, \mathbf{m}) = \boldsymbol{\epsilon} + \sigma(\boldsymbol{\rho}) \cdot \mathbf{h},
\end{equation}

\noindent $\bullet$ \textbf{Output.} After obtaining the output $\mathcal{S}(\boldsymbol{\epsilon}, \mathbf{m})$ and $\mathcal{C}(\boldsymbol{\epsilon}, \mathbf{m})$, the two are fused with a residual connection that incorporates auxiliary metadata into the output, thereby facilitating more stable training and improving model convergence. To summarize, given a randomly sampled Gaussian noise vector $\boldsymbol{\epsilon} \in \mathbb{R}^{\mathbf{D}}$ and auxiliary metadata $\mathbf{m}$, PSNet produces a refined representation $\boldsymbol{\epsilon}^{(\mathbf{m})}$. The procedure is described as follows:
\begin{equation}
\boldsymbol{\epsilon}^{(\mathbf{m})} = 
\underbrace{\mathcal{S}(\boldsymbol{\epsilon}, \mathbf{m})}_{\substack{\text{spectral} \\ \text{rectification}}}
\;+\;
\underbrace{\boldsymbol{\eta}_0 \mathcal{C}(\boldsymbol{\epsilon}, \mathbf{m})}_{\substack{\text{contextual} \\ \text{re-encoding}}}
\;+\;
\underbrace{\boldsymbol{\sigma}(\boldsymbol{\eta}_1 - 1)\, \varphi(\mathbf{m})}_{\substack{\text{residual} \\ \text{term}}},
\end{equation}

where $\sigma(\cdot)$ denotes the sigmoid function, $\varphi(\mathbf{m}) \in \mathbb{R}^{\mathbf{D}}$ is the embedding of auxiliary metadata, and $\boldsymbol{\eta}_0, \boldsymbol{\eta}_1 \in \mathbb{R}$ are learnable scalars. Through this, we obtain informative noise $\boldsymbol{\epsilon}^{(\mathbf{m})}$ enriched with semantic information, which can be employed in the subsequent diffusion process to generate embeddings that more accurately reflect genuine user preferences.

\subsection{Semantics Enhanced Contrastive Learning}
Existing contrastive recommendation systems typically generate augmented views by randomly perturbing the original interaction data. This process aims to construct sparse interaction views, thereby smoothing the distribution of item and user embeddings and enhancing the robustness. In contrast, we generate more informative embeddings as contrastive views, guiding the distribution of item representations to align more closely with user preferences. By harnessing the strong generative capacity of diffusion models to synthesize embeddings that excavate users’ latent authentic preferences, we can boost the efficacy of contrastive learning.

Specifically, the informative noise is injected into the forward process of the diffusion paradigm. For an initialized item embedding $\mathbf{e}_i \in \mathbb{R}^{\mathbf{d}}$ and informative noise $\boldsymbol{\epsilon}^{(\mathbf{m})}$ with auxiliary metadata $\mathbf{m}$, the $t$-th forward step is:
\begin{equation}
\mathbf{z}^{(\mathbf{m})}_t
= \sqrt{\bar{\alpha}_t}\, \mathbf{e}_i + \sqrt{1 - \bar{\alpha}_t}\, \boldsymbol{\epsilon}^{(\mathbf{m})}_t, \quad
\bar{\alpha}_t = \prod_{s=1}^{t} \alpha_s, \quad 0 < \alpha_t < 1.
\end{equation}
When $t = T$, $\mathbf{z}^{(\mathbf{m})}_T$ serves as the starting point for the reverse chain.

After the informative-noise-motivated diffusion, the final generated output $\hat{\mathbf{e}}^{(\mathbf{m})}_i$ thus captures both item semantics and user preferences for downstream recommendation tasks. The practical implementation of optimization for this can be expressed as:
\begin{equation}
\mathcal{L}_\mathbf{r} = \sum_{i=1}^{\mathbf{N}} \left\| \mathbf{e}_i - \boldsymbol{\mu}_{\boldsymbol{\theta}}(\mathbf{z}^{(\mathbf{m})}_t, t) \right\|^2.
\end{equation}
Here, the reconstruction loss $\mathcal{L}_\mathbf{r}$ directs the evolution of the latent state $\mathbf{z}_t$ toward true user preferences.

Subsequently, we incorporate a view-level contrastive loss that aligns the embeddings $\hat{\mathbf{e}}^{(\mathbf{m})}_i$ generated by the diffusion paradigm with the item embedding $\mathbf{e}_i$, injecting authentic latent user preferences into item embeddings, as follows:
\begin{equation}
\mathcal{L}_{\mathbf{con}} = -\sum_{i=1}^{\mathbf{N}} \log
\frac{\exp\!\left(\mathrm{sim}(\hat{\mathbf{e}}^{(\mathbf{m})}_i, \mathbf{e}_i)/\tau\right)}
     {\sum_{j=1}^{\mathbf{N}} \exp\!\left(\mathrm{sim}(\hat{\mathbf{e}}^{(\mathbf{m})}_i, \mathbf{e}_j)/\tau\right)},
\end{equation}
where $\tau > 0$ is the temperature hyperparameter.

\subsection{Collaborative Training Objective Strategy}
\label{sec:ctos}
In previous sections, we extract the semantic information in the auxiliary metadata to generate embeddings with potential user preferences through diffusion paradigm. These preferences are then integrated into the item representations by contrastive learning. Yet, as item representations are initialized from scratch, they inherently lack co‐occurrence relationships. To address this, we devise a hybrid optimization approach that integrates user–item interaction patterns into the training of the generative model. Specifically, except for the reconstruction loss $\mathcal{L}_\mathbf{r}$ and contrastive loss $\mathcal{L}_{\mathbf{con}}$, we introduce the Bayesian Personalized Ranking (BPR) loss, $\mathcal{L}_{bpr}$, as a supplementary objective, which is defined as:
\begin{equation}
    \mathcal{L}_{bpr}
    = - \sum_{u=1}^{\mathbf{M}} \sum_{i \in \mathcal{N}_{u}} \sum_{j \notin \mathcal{N}_{u}}
      \ln \sigma\bigl(\boldsymbol{\hat{y}}_{ui} - \boldsymbol{\hat{y}}_{uj}\bigr).
\end{equation}

\noindent $\bullet$ \textbf{Collaboration Balance Loss.}  
During practical training, we observe a growing gap between the reconstruction loss and the BPR loss as the model progresses, which negatively impacts the quality of generation. To improve both the stability and overall performance of our model, we introduce a \textit{collaboration balance loss}, denoted as $\mathcal{L}_{c}$, which explicitly targets the generative process. This design stems from a key insight: a significant discrepancy often exists between the latent variable $\mathbf{z}^{(\mathbf{m})}_t$ and the item representation early in training. Without appropriate constraints, this misalignment can lead to erratic or suboptimal generative behavior. Drawing inspiration from the effectiveness of regularization strategies, we employ an L2‐norm‐based formulation to regulate the generative output. The balance loss is defined as:
\begin{equation}
    \mathcal{L}_{c}
    = \frac{1}{\mathbf{N}} \bigl\| \hat{\mathbf{e}}^{(\mathbf{m})} \bigr\|_{2}
    = \frac{1}{\mathbf{N}} \Bigl( \sum_{i} \bigl| \hat{\mathbf{e}}^{(\mathbf{m})}_{i} \bigr|^{2} \Bigr)^{1/2}.
\end{equation}

\noindent $\bullet$ \textbf{Total Optimization.} Finally, the overall training objective is a weighted combination of all components:
\begin{equation}
\mathcal{L}_{total}
= (1 - \lambda_{b})\,\mathcal{L}_{r}
+ \lambda_{b}\,\mathcal{L}_{bpr}
+ \lambda_{c}\,\mathcal{L}_{con}
+ \lambda_{l}\,\mathcal{L}_{c}
+ \lambda_{g}\,\mathcal{L}_{reg},
\label{eq:loss}
\end{equation}
where the non–negative coefficients $\{\lambda_{b}, \lambda_{c}, \lambda_{l}, \lambda_{g}\}$ balance the influence of each loss term and are tuned on a validation set. The regularizer is
\begin{equation}
\mathcal{L}_{reg}
= \sum_{i=1}^{\mathbf{N}} \bigl(\|\mathbf{e}_{i}\|_{2}^{2} + \|\mathbf{u}_{i}\|_{2}^{2}\bigr),
\end{equation}
which penalizes both item embeddings $\mathbf{e}_{i}$ and user embeddings $\mathbf{u}_{i}$ to discourage representation collapse and promote smoothness.

\noindent $\bullet$ \textbf{Multiple Types of Optimization.} The above represents the total loss for a single auxiliary metadata channel. In the case of two channels, the corresponding reconstruction losses, contrastive losses, and balance losses generated by each are summed and jointly optimized as the respective components in Eq.~\ref{eq:loss}.

\subsection{Inference Stage}

During inference, we dispense with any additional diffusion sampling and directly rely on the user and item representations learned during training. To further enrich these embeddings, we propagate higher‐order co–occurrence signals using the LightGCN architecture. Concretely, we first convert the observed interaction graph into a symmetrically normalized adjacency matrix~$\bar{\boldsymbol{\mathcal{A}}}_{\mathbf{u},\mathbf{i}}$. We then perform $K$ layers of graph convolution to obtain the final user and item representations:
\begin{equation}
    \hat{\mathbf{E}} = \bar{\boldsymbol{\mathcal{A}}}_{\mathbf{u},*}\, \mathbf{E}, \qquad
    \hat{\mathbf{U}} = \bar{\boldsymbol{\mathcal{A}}}_{*,\mathbf{i}}\, \mathbf{U}, \qquad
    \bar{\boldsymbol{\mathcal{A}}}_{\mathbf{u},\mathbf{i}} =
    \frac{\boldsymbol{\mathcal{A}}_{\mathbf{u},\mathbf{i}}}
         {\sqrt{\lvert\boldsymbol{\mathcal{N}}_{\mathbf{u}}\rvert
                \lvert\boldsymbol{\mathcal{N}}_{\mathbf{i}}\rvert}}.
\end{equation}

In the above, the initial node features are set to $\mathbf{U}$ for users and $\mathbf{E}$ for items; the resulting matrices satisfy $\hat{\mathbf{U}} \in \mathbb{R}^{\mathbf{M} \times \mathbf{d}}$ and $\hat{\mathbf{E}} \in \mathbb{R}^{\mathbf{N} \times \mathbf{d}}$. The term $\boldsymbol{\mathcal{N}}_{\mathbf{u}}$ (resp.\ $\boldsymbol{\mathcal{N}}_{\mathbf{i}}$) denotes the neighbor set of user~$\mathbf{u}$ (resp.\ item~$\mathbf{i}$) in the interaction graph. Finally, we estimate a relevance score for every user–item pair by taking the inner product of their refined embeddings, and rank items by these scores to produce personalized recommendations.

\subsection{Theoretical Analysis}
To further validate the effectiveness of the proposed informative noise, we present a rigorous theoretical explanation of its generation process in Theorem \ref{theo:a}, demonstrating that the noise produced by PSNet not only conforms to the diffusion framework but also effectively incorporates semantic information from auxiliary metadata. Furthermore, in Theorem \ref{theo:b}, we undertake an in-depth analysis of the impact of informative noise on user preference learning, showing that it guides the diffusion process to yield representations with higher user preference scores.

\begin{table}[t]
    \centering
    \normalsize
    \captionsetup{skip=0pt}
    \caption{Statistics of the datasets}
    \setlength{\tabcolsep}{3pt}
    \begin{tabular}{lrrrrr}
        \toprule
        \textbf{Datasets} & \textbf{ML-1M} & \textbf{Office} & \textbf{Baby} & \textbf{Taobao} & \textbf{Electronics}\\
        \midrule
        \textbf{\#Users} & 6040 & 4,905 & 19,445 & 12,539 & 32,886\\
        \textbf{\#Items} & 3706 & 2,420 & 7,050 & 8,735 & 52,974\\
        \textbf{\#Int.} & 1,000,209 & 53,258 & 159,669 & 83,648 & 337,837\\
        \textbf{Sparsity} & 95.53\% & 99.55\% & 99.88\% & 99.92\% & 99.69\%\\
        \bottomrule
    \end{tabular}
    \label{tab:data}
\vspace{-0.1in}
\end{table}
\section{Experiment}
\subsection{Experimental Settings}
\subsubsection{\textbf{Datasets}}
\  
\newline
Experimental evaluations are conducted on five commonly used public recommendation datasets, including ml-1m, Amazon-Office, Amazon-Electronics, Amazon-Baby, and Taobao. The statistics of each dataset are listed in Table \ref{tab:data}.
\subsubsection{\textbf{Evaluation Metrics}}
\ 
\newline
The effectiveness of the proposed recommender system is assessed using two widely adopted ranking metrics: \textbf{NDCG@\textit{K}} and \textbf{Recall@\textit{K}}, with \textit{K} indicating the cutoff position in the ranked list. The all-ranking evaluation protocol is employed, and the final performance is reported as the average score over all test users.

\subsubsection{\textbf{Baseline Models}}
\ 
\newline
A comprehensive performance comparison is conducted between our proposed framework, InfoDCL, and a broad set of baseline methods. These baselines include: (1) classical collaborative filtering models such as Matrix Factorization (\textbf{MF}\cite{mf}) and \textbf{ENMF}\cite{enmf}; (2) representative graph neural network-based approaches, including \textbf{NGCF}\cite{ngcf} and \textbf{LightGCN}\cite{lightgcn}; (3) state-of-the-art generative models based on diffusion processes, including \textbf{DiffRec}\cite{diffrec}, \textbf{DDRM}\cite{ddrm} and \textbf{GiffCF}\cite{giffcf}; and (4) recent contrastive learning-based techniques with strong accuracy, such as \textbf{SGL}\cite{sgl}, \textbf{NCL}\cite{ncl}, \textbf{SimGCL}\cite{simgcl}, \textbf{RecDCL}\cite{recdcl}, \textbf{SGCL}\cite{sgcl} and \textbf{CoGCL}\cite{cogcl}.

\subsubsection{\textbf{Implementation Details}}
\ 
\newline
All models are configured with a consistent embedding dimensionality of 64, initialized using the Xavier scheme. The hyperparameter optimization process spans several key dimensions. Specifically, the learning rate is logarithmically sampled within the interval $[1 \times 10^{-6},\, 5 \times 10^{-1}]$. Batch sizes are discretely selected based on the interaction density of each dataset to maintain training efficiency---for example, a batch size of 1024 is adopted for the \textit{ML-1M} dataset, while 2000 is used for \textit{Amazon-Office}. The loss coefficients governing the diffusion reconstruction term ($\lambda_{r}$) and pairwise ranking term ($\lambda_{b}$) are both tuned within the range $[0,\,1.0]$, whereas the regularization coefficient ($\lambda_{g}$) is searched in the interval $[0.001,\,0.01]$ to ensure adequate generalization. In addition, the contrastive loss weight ($\lambda_{\text{con}}$) is varied between $5 \times 10^{-5}$ and $5 \times 10^{-6}$. The number of GCN layers used during inference is explored from 0 to 3, and the number of diffusion timesteps is tested across a range of 100 to 500. Finally, we compare the impact of using \textit{Adam} and \textit{AdamW} optimizers, both of which are well-established in the deep learning community for their effectiveness and stability.
\begin{table*}[!ht]
\centering
\scriptsize
\setlength{\tabcolsep}{4pt}  % ← 缩小列间距
\renewcommand{\arraystretch}{1.3}  % ← 增加行间距
\caption{Performance comparison of different methods on the five datasets. The best and second-best performances are indicated in bold and underlined font, respectively. InfoDCL-S and InfoDCL-D denote the use of single-channel and dual-channel contrastive learning within our framework.}
\vspace{-0.1in}
\begin{tabular}{c|c|cc|cc|ccc|cccccc|cc|c}
\hline
\multirow{2}{*}{Dataset} & \multirow{2}{*}{Metric} 
& \multicolumn{2}{c|}{Matrix Factorization} 
& \multicolumn{2}{c|}{GCN-based Models} 
& \multicolumn{3}{c|}{Diffusion Models} 
& \multicolumn{6}{c|}{Contrastive Learning Models} 
& \multicolumn{2}{c|}{Ours} 
& \multirow{2}{*}{\textit{Improve.}} \\
\cline{3-17}
& & MF & ENMF & NGCF & LightGCN & DiffRec & DDRM & GiffCF & SGL & NCL & SimGCL & RecDCL & SGCL & CoGCL & InfoDCL-S & InfoDCL-D & \\
\hline
\multirow{4}{*}{\centering Baby} 
 & Recall@20 & 0.0451 & 0.0602 & 0.0532 & 0.0715 & 0.0713 & 0.0118 & 0.0725 & 0.0656 & 0.0742 & \underline{0.0782} & 0.0726 & 0.0533 & 0.0765 & 0.0832 & \textbf{0.0856} & 9.46\% \\
 & Recall@50 & 0.0899 & 0.1055 & 0.1002 & 0.1255 & 0.1181 & 0.0178 & 0.1253 & 0.1090 & 0.1305 & \underline{0.1324} & 0.1142 & 0.0839 & 0.1289 & 0.01417 & \textbf{0.1445} & 9.14\% \\
 & NDCG@20 & 0.0185 & 0.0287 & 0.0226 & 0.0298 & 0.0327 & 0.0051 & 0.0323 & 0.0297 & 0.0321 & 0.0332 & \underline{0.0338} & 0.0240 & 0.0318 & 0.0355 & \textbf{0.0359} & 6.21\% \\
 & NDCG@50 & 0.0272 & 0.0377 & 0.0320 & 0.0409 & 0.0422 & 0.0063 & \underline{0.0449} & 0.0384 & 0.0433 & 0.0443 & 0.0431 & 0.0301 & 0.0427 & 0.0472 & \textbf{0.0477} & 6.29\% \\
\hline
\multirow{4}{*}{\centering Office} 
 & Recall@20 & 0.0598 & 0.1004 & 0.0928 & 0.1215 & 0.1159 & 0.0133 & 0.1252 & 0.1151 & 0.0966 & \underline{0.1305} & 0.1254 & 0.0414 & 0.1206 & 0.01364 & \textbf{0.1398} & 11.70\% \\
 & Recall@50 & 0.1178 & 0.1729 & 0.1684 & 0.2064 & 0.1867 & 0.0277 & \underline{0.2084} & 0.1838 & 0.1595 & 0.2073 & 0.1969 & 0.0705 & 0.1986 & 0.2217 & \textbf{0.2282} & 9.48\% \\
 & NDCG@20 & 0.0232 & 0.0500 & 0.0400 & 0.0558 & 0.0511 & 0.0058 & 0.0537 & 0.0549 & 0.0463 & 0.0562 & 0.0534 & 0.0206 & \underline{0.0572} & \textbf{0.0630} & 0.0625 & 10.14\% \\
 & NDCG@50 & 0.0346 & 0.0651 & 0.0563 & 0.0702 & 0.0704 & 0.0088 & 0.0719 & 0.0697 & 0.0594 & 0.0733 & 0.0689 & 0.0267 & \underline{0.0737} & 0.0810 & \textbf{0.0812} & 10.18\% \\
\hline
\multirow{4}{*}{\centering Taobao} 
 & Recall@20 & 0.0556 & 0.1307 & 0.1223 & 0.1502 & 0.1492 & 0.0139 & 0.1524 & 0.1555 & 0.1558 & \underline{0.1611} & 0.1459 & 0.1334 & 0.1458 & \textbf{0.2001} & 0.1996 & 24.20\% \\
 & Recall@50 & 0.0983 & 0.1813 & 0.1902 & 0.2250 & 0.2013 & 0.0228 & 0.2084 & 0.2107 & \underline{0.2372} & 0.2189 & 0.2114 & 0.1890 & 0.1957 & \textbf{0.2837} & 0.2825 & 19.60\% \\
 & NDCG@20 & 0.0207 & 0.0630 & 0.0523 & 0.0681 & 0.0715 & 0.0057 & 0.0659 & 0.0748 & 0.0717 & \underline{0.0762} & 0.0713 & 0.0634 & 0.0720 & \textbf{0.0900} & 0.0895 & 18.11\% \\
 & NDCG@50 & 0.0290 & 0.0731 & 0.0658 & 0.0830 & 0.0824 & 0.0075 & 0.0786 & 0.0859 & 0.0880 & \underline{0.0898} & 0.0844 & 0.0746 & 0.0820 & \textbf{0.1066} & 0.1060 & 18.71\% \\
\hline
\multirow{4}{*}{\centering Electronics} 
 & Recall@20 & 0.0401 & 0.0299 & 0.0368 & 0.0394 & 0.0236 & 0.0033 & 0.0343 & 0.0359 & \underline{0.0435} & 0.0423 & 0.0409 & 0.0407 & 0.0415 & \textbf{0.0473} & 0.0464 & 8.73\% \\
 & Recall@50 & 0.0620 & 0.0512 & 0.0593 & 0.0645 & 0.0451 & 0.0044 & 0.0509 & 0.0561 & \underline{0.0679} & 0.0655 & 0.0614 & 0.0622 & 0.0648 & 0.0715 & \textbf{0.0735} & 8.25\% \\
 & NDCG@20 & 0.0155 & 0.0139 & 0.0163 & 0.0178 & 0.0123 & 0.0020 & 0.0138 & 0.0175 & 0.0199 & \underline{0.0192} & 0.0182 & 0.0198 & 0.0192 & 0.0210 & \textbf{0.0217} & 9.05\% \\
 & NDCG@50 & 0.0198 & 0.0183 & 0.0209 & 0.0229 & 0.0189 & 0.0022 & 0.0181 & 0.0217 & \underline{0.0249} & 0.0230 & 0.0213 & 0.0243 & 0.0239 & 0.0261 & \textbf{0.0273} & 9.64\% \\
\hline
\multirow{4}{*}{\centering ML-1M} 
 & Recall@20 & 0.0751 & 0.1061 & 0.0877 & 0.0790 & 0.0794 & 0.0141 & 0.1044 & 0.0778 & 0.0868 & \underline{0.1192} & 0.0858 & 0.0107 & 0.1020 & \textbf{0.1701} & 0.1631 & 42.70\% \\
 & Recall@50 & 0.0854 & 0.2154 & 0.1785 & 0.1666 & 0.1761 & 0.0182 & 0.2006 & 0.1719 & 0.1740 & \underline{0.2167} & 0.1709 & 0.0211 & 0.1977 & \textbf{0.2505} & 0.2450 & 15.59\% \\
 & NDCG@20 & 0.0244 & 0.0402 & 0.0347 & 0.0278 & 0.0316 & 0.0059 & 0.0379 & 0.0304 & 0.0310 & \underline{0.0417} & 0.0321 & 0.0123 & 0.0371 & \textbf{0.0538} & 0.0511 & 29.02\% \\
 & NDCG@50 & 0.0263 & 0.0554 & 0.0525 & 0.0451 & 0.0468 & 0.0068 & 0.0569 & 0.0489 & 0.0480 & \underline{0.0581} & 0.0487 & 0.0155 & 0.0560 & \textbf{0.0696} & 0.0673 & 19.79\% \\
\hline
\end{tabular}
\label{tab:performance}
\end{table*}
\begin{figure*}
    \centering
    \captionsetup{skip=0pt}
    \includegraphics[width=\textwidth]{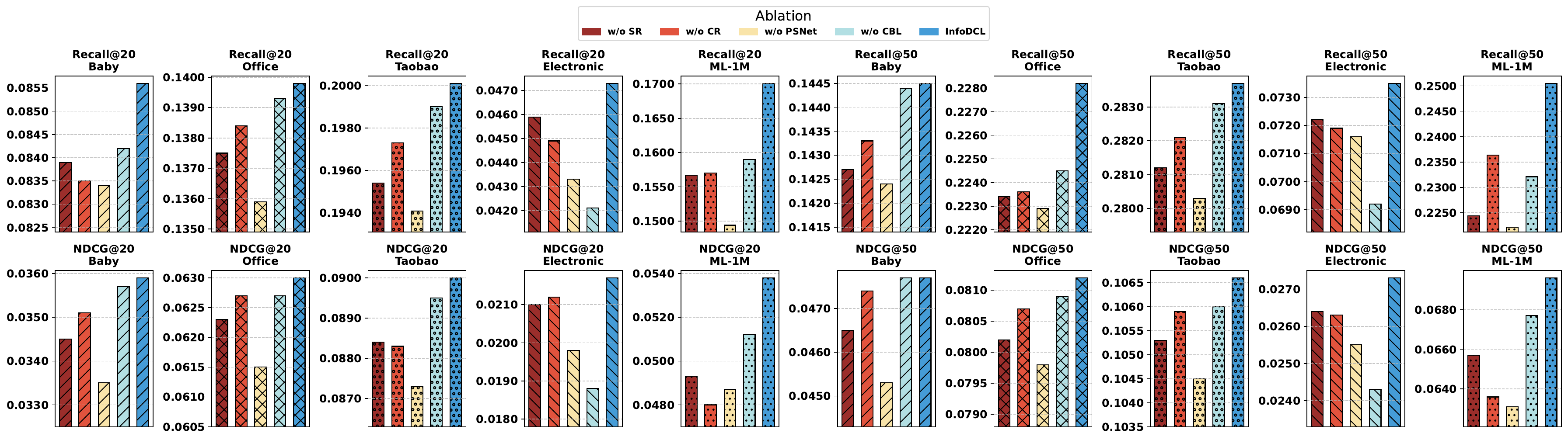}
    \caption{Ablation analysis across five datasets}
    \label{fig:ab}
\end{figure*}
\vspace{-0.1in}
\subsection{Performance Comparison}
Table \ref{tab:performance} presents a comparative analysis of our proposed model against various baseline models across five datasets, from which we can have following observations:
\begin{itemize}[leftmargin=*]
    \item Diffusion-based recommender systems, such as DiffRec and GiffCF, achieve superior performance over traditional baselines by leveraging noise-injection mechanisms and reverse denoising processes to capture intricate dependencies between user and item interactions. Their generative framework offers more diverse recommendations, making it easier for users to explore different kinds of content. However, the randomly sampled noise corrupts the already sparse user-item interactions, decreasing the generative ability of diffusion model.
    \item Contrastive learning methods, including SGL and CoGCL, which generate positive and negative pairs, encouraging consistency of the same node across augmented views while maintaining representational distinctiveness among different nodes. However, these models typically apply random perturbations to existing interactions to create new contrastive views, which drives the embeddings toward uniformity and thereby enhances robustness. They nevertheless fail to leverage supplementary information, which constrains their predictive performance.
    \item Our InfoDCL outperforms other state-of-the-art models in metrics across all datasets, achieving the best overall performance. This highlights the value of leveraging auxiliary metadata and SVD to produce informative noise, which enhances the generation of embeddings in the diffusion process with semantic information in contrast to random Gaussian noise. Additionally, the collaborative training objective strategy transforms the mutual interference between generation, contrastive learning and preference learning into collaboration, significantly improving predictive performance.
\end{itemize}

\subsection{Ablation Analysis}

Figure \ref{fig:ab} presents the ablation study results. In this analysis, \textbf{w/o SR} denotes the removal of the spectral rectification module, leaving only the contextual re-encoding to integrate Gaussian noise with auxiliary metadata. In other words, this configuration discards the use of singular value decomposition (SVD) to simulate a one-step diffusion process for generating semantically rich noise, resulting in a simple linear fusion of noise and semantics. As illustrated in Figure \ref{fig:ab}, eliminating the spectral rectification module leads to a significant performance drop.

Similarly, \textbf{w/o CR} refers to the removal of the contextual re-encoding module, using only SVD to generate informative noise. Although its performance degradation is less severe than that observed in \textbf{w/o SR}, the decline still highlights the insufficiency of relying solely on SVD to approximate the one-step diffusion process. This underscores the necessity for additional enhancements, and our proposed contextual re-encoding module provides an effective solution.

Additionally, we designed the variant \textbf{w/o PSNet}, which replaces the informative noise with randomly sampled Gaussian noise, corresponding to the standard diffusion process for generating item embeddings. As shown in the Figure \ref{fig:ab}, across almost all datasets and evaluation metrics, the removal of PSNet leads to a severe performance degradation. This result strongly indicates that the informative noise, generated through the injection of auxiliary semantic information, effectively leverages the diffusion paradigm to produce item embeddings that better capture authentic user preferences.
\begin{table}[!t]
\centering
\scriptsize
\setlength{\tabcolsep}{2pt}
\renewcommand{\arraystretch}{1}
\caption{Performance comparison of different multimodal methods on the five datasets.}
\begin{tabular}{c|c|cccccc|>{\bfseries}c}
\hline
Datasets & Metric & MMSSL & LATTICE & BM3 & LGMRec & MGCN & DiffMM & \textbf{InfoDCL}\\
\hline
\multirow{4}{*}{Baby}
 & Recall@20 & 0.0525 & 0.0839 & 0.0835 & 0.0643 & 0.0744 & 0.0806 & 0.0890 \\
 & Recall@50 & 0.1005 & 0.1432 & 0.1456 & 0.1157 & 0.1326 & 0.1459 & 0.1476 \\
 & NDCG@20 & 0.0228 & 0.0368 & 0.0364 & 0.0284 & 0.0327 & 0.0337 & 0.0375 \\
 & NDCG@50 & 0.0327 & 0.0473 & 0.0465 & 0.0388 & 0.0455 & 0.0463 & 0.0492 \\
\hline
\multirow{4}{*}{Office}
 & Recall@20 & 0.1277 & 0.1345 & 0.1158 & 0.1348 & 0.1196 & 0.1351 & 0.1438 \\
 & Recall@50 & 0.2123 & 0.2200 & 0.1944 & 0.2231 & 0.2029 & 0.2308 & 0.2280 \\
 & NDCG@20 & 0.0541 & 0.0524 & 0.0527 & 0.0598 & 0.0544 & 0.0599 & 0.0644 \\
 & NDCG@50 & 0.0732 & 0.0742 & 0.0695 & 0.0789 & 0.0724 & 0.0804 & 0.0823 \\
\hline
\multirow{4}{*}{Taobao}
 & Recall@20 & 0.1619 & 0.1622 & 0.1451 & 0.1661 & 0.1528 & 0.1498 & 0.2006 \\
 & Recall@50 & 0.2377 & 0.2434 & 0.2246 & 0.2392 & 0.2411 & 0.2342 & 0.2818 \\
 & NDCG@20 & 0.0749 & 0.0699 & 0.0636 & 0.0693 & 0.0645 & 0.0649 & 0.0904 \\
 & NDCG@50 & 0.0901 & 0.0862 & 0.0802 & 0.0868 & 0.0829 & 0.0817 & 0.1066 \\
\hline
\multirow{4}{*}{Electronics}
 & Recall@20 & 0.0425 & 0.0461 & 0.0451 & 0.0449 & 0.0466 & 0.0467 & 0.0533 \\
 & Recall@50 & 0.0671 & 0.0712 & 0.0738 & 0.0733 & 0.0756 & 0.0754 & 0.0851 \\
 & NDCG@20 & 0.0214 & 0.0206 & 0.0207 & 0.0209 & 0.0212 & 0.0215 & 0.0244 \\
 & NDCG@50 & 0.0273 & 0.0264 & 0.0267 & 0.0268 & 0.0274 & 0.0277 & 0.0310 \\
\hline
\multirow{4}{*}{ML-1M}
 & Recall@20 & 0.0683 & 0.0743 & 0.1005 & 0.1507 & 0.0844 & 0.0854 & 0.1647 \\
 & Recall@50 & 0.1474 & 0.1745 & 0.1983 & 0.2369 & 0.1826 & 0.1895 & 0.2494 \\
 & NDCG@20 & 0.0242 & 0.0271 & 0.0319 & 0.0412 & 0.0332 & 0.0336 & 0.0462 \\
 & NDCG@50 & 0.0397 & 0.0467 & 0.0512 & 0.0569 & 0.0517 & 0.544 & 0.0630 \\
\hline
\end{tabular}
\label{tab:mmssl_comparison}
\end{table}

Furthermore, the variant \textbf{w/o CBL} denotes removing the collaboration balance loss from the overall optimization, which is introduced in our collaborative training objective strategy. As shown in Figure \ref{fig:ab}, this loss term is equally crucial to model performance. First, omitting this loss leads to performance degradation across all datasets and evaluation metrics. Second, on certain datasets, its impact on performance enhancement is especially significant. For example, on Amazon-Electronics, the \textbf{w/o CBL} variant shows the most pronounced decline. This highlights the important role of the collaboration balance loss in coordinating the multiple learning objectives.

\subsection{Comparison with Multimodal Baselines}

To further evaluate the effectiveness of InfoDCL in leveraging modality information, we conduct experiments to compare its performance with several state-of-the-art multimodal recommender systems, including \textbf{MMSSL}\cite{mmssl}, \textbf{LATTICE}\cite{lattice}, \textbf{BM3}\cite{bm3}, \textbf{LGMRec}\cite{lgmrec}, \textbf{MGCN}\cite{mgcn}, and \textbf{DiffMM}\cite{DiffMM}. All these baselines as well as InfoDCL, incorporate both visual and textual modalities, except on the Taobao dataset, which contains only visual data. Across the five datasets, InfoDCL consistently achieves superior performance compared to these baselines.

The results suggest that InfoDCL is more effective at utilizing modality information than existing multimodal models. Notably, on the Taobao dataset, where only a single modality is present, InfoDCL significantly outperforms all competitors. This finding indicates that the proposed PSNet certainly generates informative noise with semantic information, which enables the diffusion process to better capture users' genuine preferences. It also demonstrates that the generated item embeddings inject rich semantic information to the initialized item representation through the collaborative training objective strategy, further improving recommendation performance.
\subsection{Training Efficiency}

In this section, we assess the trade-off between model performance and training efficiency by comparing a range of representative recommendation models on the Amazon-Electronics dataset, which contains the largest number of users and items among all evaluated datasets. To provide a fair and consistent evaluation, we report two metrics for each model: the average training time per epoch and the Recall@20. All experiments are conducted using the same GPU under a single-process setting to eliminate hardware and implementation variability.
\begin{figure}
    \centering
    \captionsetup{skip=0pt}
    \includegraphics[width=0.4\textwidth]{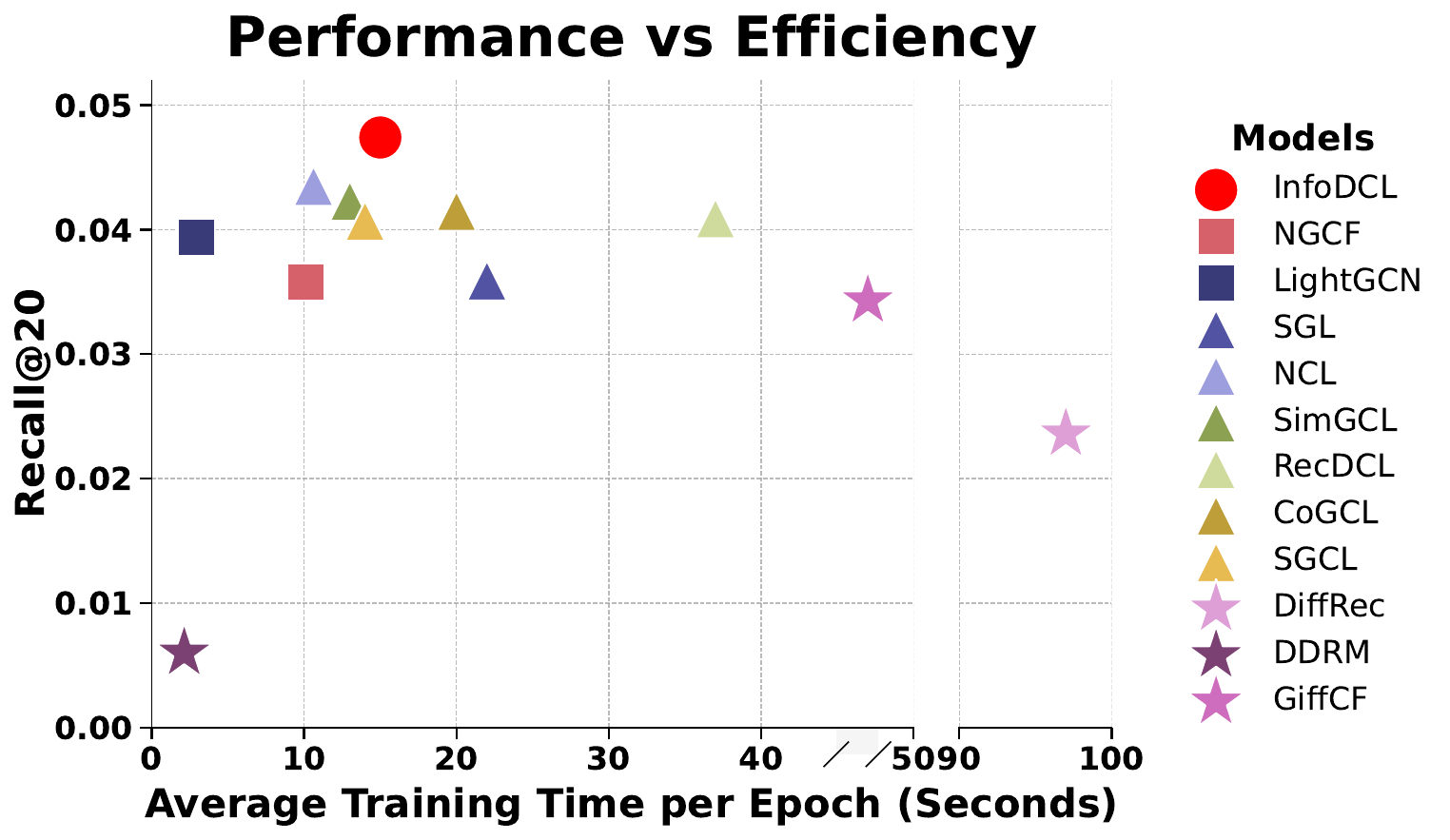}
    \caption{Performance versus efficiency analysis on Amazon-Electronics.}
    \label{fig:eff}
\end{figure}
\begin{figure}
    \centering
    \captionsetup{skip=0pt}
    \includegraphics[width=0.45\textwidth]{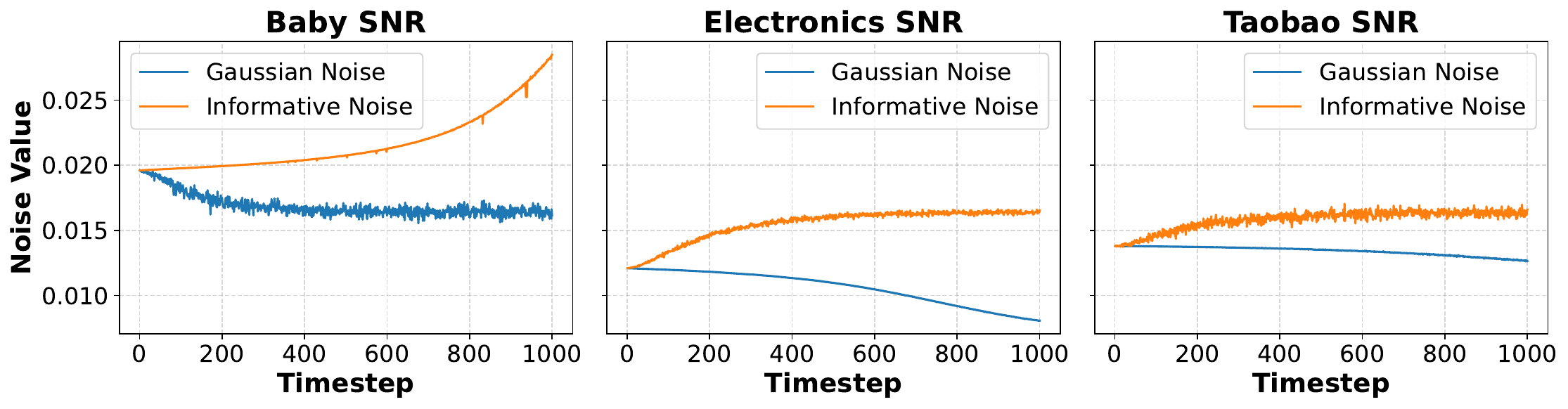}
    \caption{SNR Comparison on Amazon-Baby, Amazon-Electronics and Taobao}
    \label{fig:snr}
\end{figure}

As shown in Figure~\ref{fig:eff}, the results reveal diverse performance-efficiency trade-offs across different modeling paradigms. Among GCN-based methods, \textbf{LightGCN} stands out for its simplicity and effectiveness. By discarding non-linear transformations and feature aggregation, it maintains a lightweight architecture that achieves both accuracy and training efficiency. In contrast, diffusion-based models such as \textbf{DiffRec} attempt to leverage the dynamics of information propagation, but their increased complexity leads to significantly longer training times with only marginal gains in performance. Meanwhile, contrastive learning approaches like \textbf{NCL} demonstrate strong performance, achieving the second-highest Recall@20, although at the cost of slightly increased training time compared to LightGCN. This reflects the effectiveness of contrastive objectives in learning robust representations for recommendation.

Most notably, our proposed method, \textbf{InfoDCL}, achieves a superior balance between training efficiency and model performance. It consistently outperforms all baselines in terms of Recall@20, while keeping the training time per epoch relatively low. This optimal position in the performance-efficiency space highlights InfoDCL's ability to extract informative representations without incurring substantial computational cost.

\subsection{In-depth Analysis}
\subsubsection{\textbf{SNR Comparison Between Gaussian Noise and Informative Noise}}
\
\newline
To further evaluate the strength of our informative noise, we conduct a comparative analysis of the signal-to-noise ratio (SNR) for three benchmark datasets: \textbf{Amazon-Electronics}, \textbf{Amazon-Baby} and \textbf{Taobao}. The SNR is computed using a standard statistical definition, which denotes the ratio between the square of the mean and the variance of a random variable across time steps of noise addition. Specifically, for any random variable \( \mathbf{X} \), the SNR is given by:
\begin{equation}
\text{SNR}(\mathbf{X}) = \frac{(\mathbb{E}[\mathbf{X}])^2}{\text{Var}(\mathbf{X})},
\end{equation}
where \( \mathbb{E}[\mathbf{X}] \) denotes the expectation and \( \text{Var}(\mathbf{X}) \) represents the variance of \( \mathbf{X} \).
\begin{figure}
    \centering
    \captionsetup{skip=0pt}
    \includegraphics[width=0.45\textwidth]{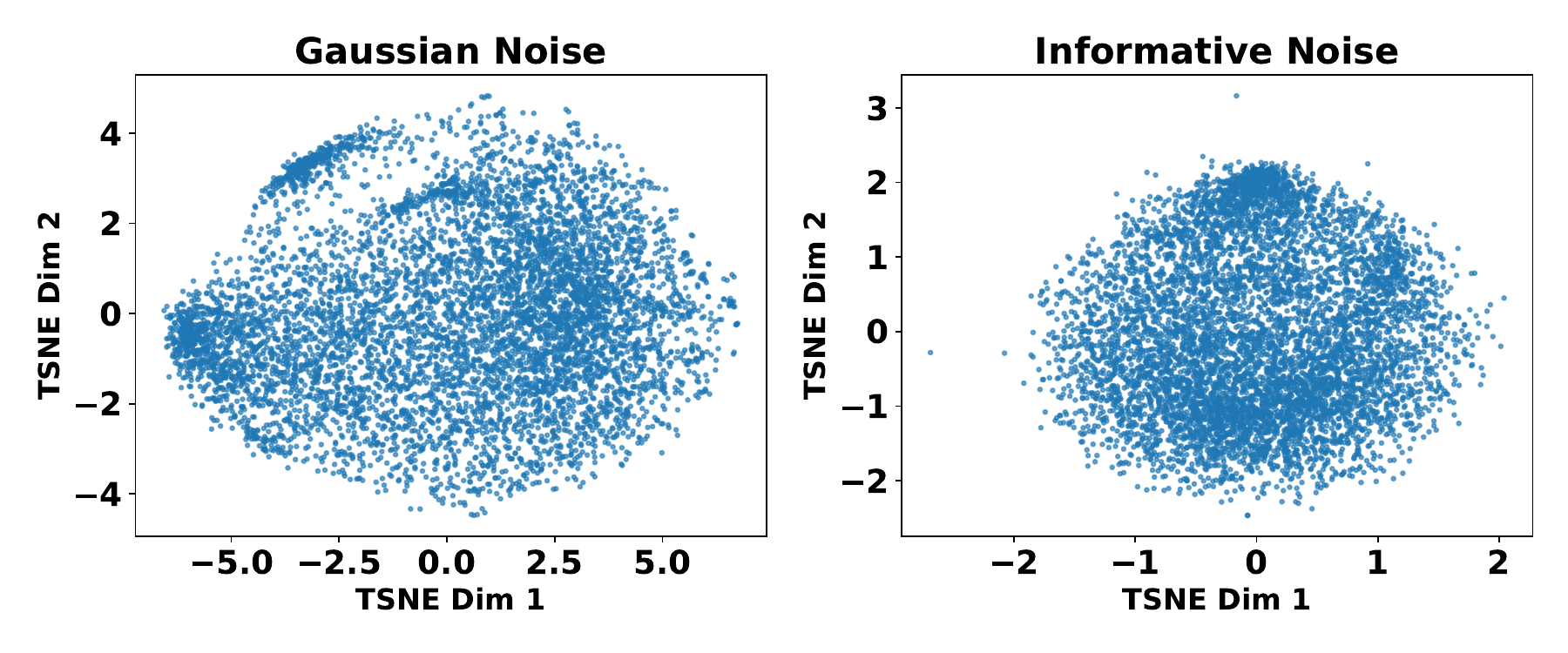}
    \caption{Visualization of the item embeddings on Amazon-Baby dataset using T-SNE.}
    \label{fig:tsne}
\end{figure}
\begin{figure}
    \centering
    \captionsetup{skip=0pt}
    \includegraphics[width=0.45\textwidth]{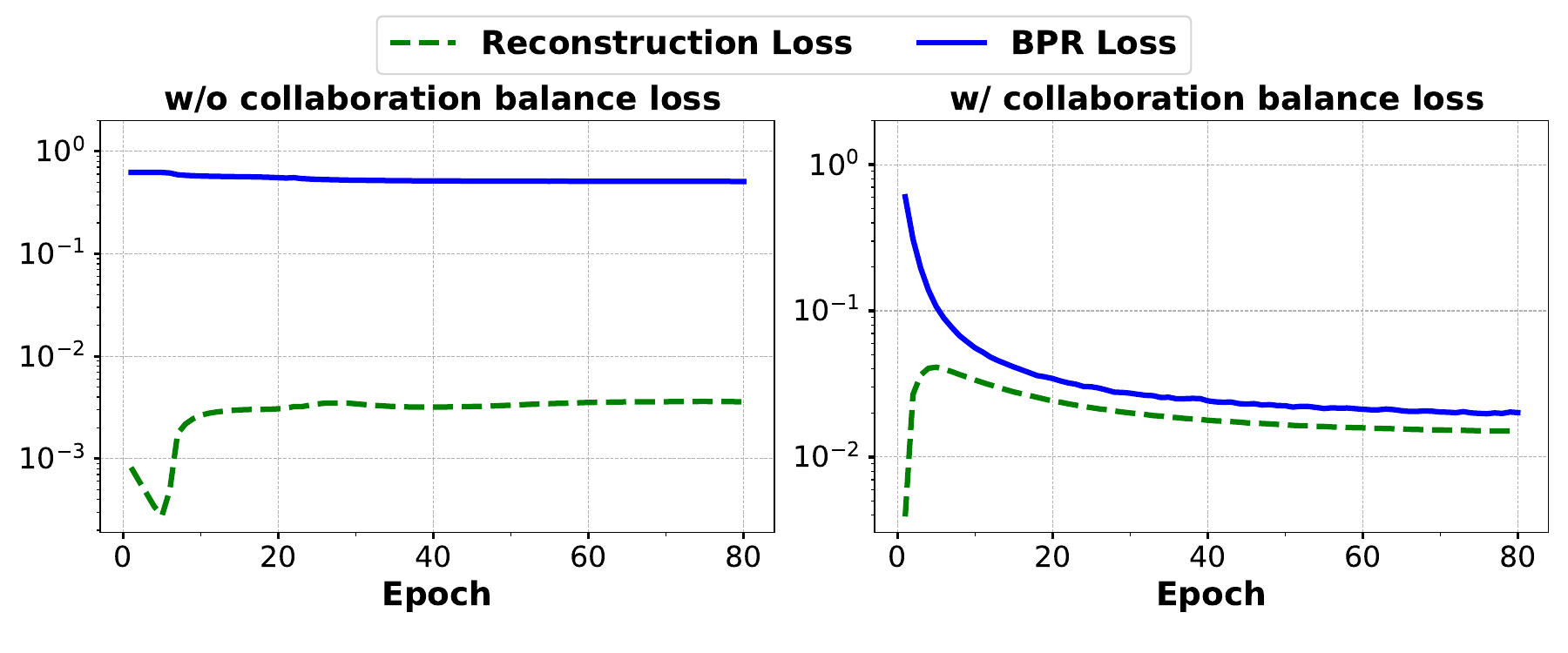}
    \caption{Comparison of w/o collaboration balance loss and InfoDCL.}
    \label{fig:loss}
\end{figure}

To capture the temporal evolution of signal informativeness, we plot the SNR values of latent variable $\mathbf{z^{(m)}_T}$ throughout time steps for both datasets, as illustrated in Figure~\ref{fig:snr}. The results demonstrate that the SNR of $\mathbf{z^{(m)}_T}$ with added informative noise consistently surpasses that of randomly sampled Gaussian noise, with the advantage becoming more pronounced as the number of noise-injection time steps increases.

This finding highlights the efficacy of our proposed informative noise generation process. By embedding rich semantic information into Gaussian noise, we are able to synthesize semantically structured noise that contains more meaningful information. This enhanced noise not only preserves the nature of the input but also provides informative guidance during the diffusion process. Consequently, the generated item embeddings more accurately capture user preferences, leading to improved representation learning and recommendation performance.
\subsubsection{\textbf{Visualization Comparison Between Gaussian Noise and Informative Noise}}
\
\newline
In order to investigate the corruption of co-occurrence relationships in recommendation data caused by randomly sampled Gaussian noise, as well as validating the effectiveness of informative noise proposed in our InfoDCL, we collect two kinds of item representations on Amazon-Baby dataset by training InfoDCL and its variant which replaces the informative noise with Gaussian noise, then we visualize them using t-SNE for intuitive observation of data distributions.

As shown in Figure \ref{fig:tsne}, the item embeddings by the variant which uses random noise demonstrate crowding in limited discrete regions of the item space, making them indistinguishable. In stark contrast, after informative noise injection in InfoDCL, the generated embeddings exhibit a more balanced spatial arrangement. This empirical observation strongly suggests that introducing noise to inherently sparse recommendation data significantly disrupts the original interaction patterns. Conversely, introducing informative noise can enrich the semantic content of the generated embeddings, enhance the modeling of user preferences, and thus be more suitable for recommendation scenarios.
\subsubsection{\textbf{Analysis of Collaborative Training Objective Strategy}}
\
\newline
\vspace{0.05in}
\begin{figure}
    \centering
    \captionsetup{skip=0pt}
    \includegraphics[width=0.4\textwidth]{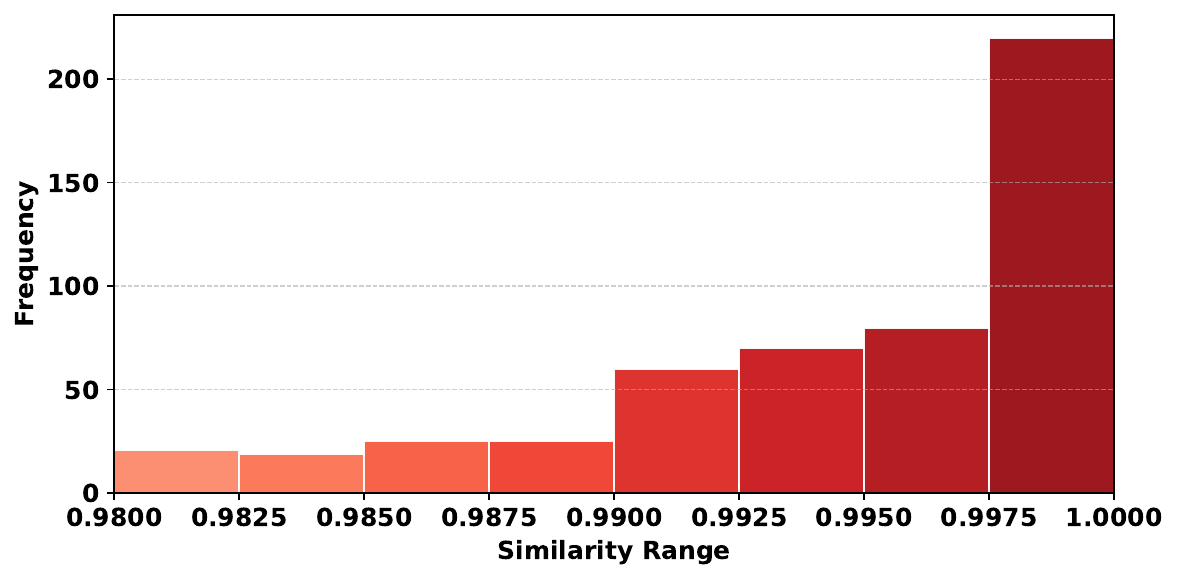}
    \caption{Cosine similarity between the singular vectors of input and output in diffusion process}
    \label{fig:svd}
\end{figure}
In this subsection, we present the core idea behind our collaborative training objective approach. Similar to other diffusion-based frameworks, InfoDCL’s reconstruction loss converges rapidly during the training process. As elaborated in Section \ref{sec:ctos}, we employ a collaborative strategy that jointly optimizes both the generation process and the recommendation objectives. Figure \ref{fig:loss} illustrates that, on the Amazon-Electronics dataset, a clear divergence arises between the reconstruction loss and the BPR loss as training continues. Eventually, the reconstruction loss diminishes to a point where its contribution to the overall loss becomes negligible, hindering further refinement of the denoising generative model. However, the introduction of our collaboration balance loss visibly improves this situation. Prior to applying this strategy, the two loss terms differed significantly in scale. Once integrated, the loss terms align more closely in magnitude, enabling consistent and stable training. This balanced optimization not only stabilizes the generative training but also leads to notable improvements in model effectiveness, as supported by our experimental outcomes.
\subsubsection{\textbf{Visualization of Similarity Between Eigenvectors}}
\vspace{0.03in}
\
\newline
To demonstrate that the singular vectors obtained from the SVD decomposition of the diffusion model’s inputs and outputs exhibit a high degree of similarity, we compute the absolute value of their cosine similarity, as shown in Figure \ref{fig:svd}. As the figure illustrates, most vector pairs attain similarity scores very close to 1, indicating strong alignment in their feature representations. This finding provides solid theoretical justification for our PSNet design, as it confirms the feasibility of transforming characteristics between the input and output spaces.
\section{Conclusion}
In this work, we propose a novel diffusion based contrastive learning framework called InfoDCL. We replace the randomly sampled Gaussian noise by informative noise, which introduces rich semantic information to sparse data. In order to improve the generative effect, we build a collaborative training objective strategy which harmonizes the generation and preference learning. Our empirical evaluations across five real-world datasets show that InfoDCL significantly outperforms previous diffusion methods. This work presents a novel perspective on contrastive learning recommender systems and suggests new research directions for applying the diffusion paradigm to inherently sparse recommendation tasks.

% \begin{acks}
% This work was supported by the National Natural Science Foundation of China under Grants 62402377, U2441242, and the Shaanxi Provincial Sanqin Talent Introduction Program for Young Talents under Grants 2024SYJ13.
% \end{acks}

\bibliographystyle{ACM-Reference-Format}
\bibliography{reference}

\appendix

\section{Theoretical Analysis}
\label{ap:A}
We provide a closed-form expression and rigorous derivation for the informative noise constructed by one round of re-denoising in the latent space of our model. Let $\mathbf{v}_0 \in \mathbb{R}^{\mathbf{d}}$ denote the clean embedding of an item, and let the forward diffusion process at any time $t$ follow the standard DDPM formulation:
\begin{equation}
\mathbf{v}_t = \boldsymbol{\alpha}_t \mathbf{v}_0 + \boldsymbol{\sigma}_t \boldsymbol{\varepsilon}, \qquad \boldsymbol{\varepsilon} \sim \mathcal{N}(\mathbf{0}, \mathbf{I}_d),
\end{equation}
where $\{ \boldsymbol{\alpha}_t, \boldsymbol{\sigma}_t \}_{t=0}^{T}$ are fixed scheduler parameters. Let $\mathbf{v}_T \sim \mathcal{N}(\mathbf{0}, \mathbf{I}_d)$ be a standard Gaussian sample serving as the initial state of the reverse process. We aim to inject structural semantic information $\mathbf{s} \in \mathbb{R}^{\mathbf{d}_s}$ --- derived from knowledge graphs or social networks --- into the noise prior to actual denoising.

\vspace{1em}
\begin{theorem}\label{theo:a}
\textit{Let $\boldsymbol{\varepsilon}_{\boldsymbol{\theta}}(\mathbf{v}, t \mid \star)$ be a trained noise prediction network that is $L$-Lipschitz with respect to its first argument.  
Assume the trajectory is smooth over a step size $k \ll T$, i.e.,}
\begin{equation}
\left\| \mathbf{v}_T - \mathbf{v}_{T-k} \right\| \leq Lk.
\end{equation}
\textit{Let the classifier-free guidance (CFG) scale at step $T$ and $T-k$ be $\boldsymbol{\omega}_{\ell}$ and $\boldsymbol{\omega}_{w}$, respectively. Then the output of one round of re-denoising, denoted $\mathbf{v}'_T$, satisfies:}
\begin{equation}
\mathbf{v}'_T = \mathbf{v}_T + \boldsymbol{\kappa} \cdot \mathbf{g}_s, \qquad
\boldsymbol{\kappa} = \left( \boldsymbol{\omega}_{\ell} - \boldsymbol{\omega}_{w} \right) \cdot 
\frac{ \boldsymbol{\alpha}_{T} \boldsymbol{\sigma}_{T-k} - \boldsymbol{\alpha}_{T-k} \boldsymbol{\sigma}_{T} }{ \boldsymbol{\alpha}_{T-k} },
\label{eq:simple}
\end{equation}
\textit{where}
\begin{equation}
\mathbf{g}_s = \boldsymbol{\varepsilon}_{\boldsymbol{\theta}} \left( \mathbf{v}_{T - \frac{k}{2}}, T - \tfrac{k}{2} \mid \mathbf{s} \right)
            - \boldsymbol{\varepsilon}_{\boldsymbol{\theta}} \left( \mathbf{v}_{T - \frac{k}{2}}, T - \tfrac{k}{2} \mid \varnothing \right)
\end{equation}
\textit{is referred to as the \textit{semantic gradient}.}
\end{theorem}
\vspace{1em}
\noindent \textit{Proof.}  
We first compute the result of one DDIM step from $T$ to $T-k$, using the deterministic update:
\begin{equation}
\mathbf{v}_{T-k} = \frac{\boldsymbol{\alpha}_{T-k}}{\boldsymbol{\alpha}_T} \left( \mathbf{v}_T - \boldsymbol{\sigma}_T \boldsymbol{\varepsilon}_{\boldsymbol{\theta}}(\mathbf{v}_T, T) \right)
+ \boldsymbol{\sigma}_{T-k} \boldsymbol{\varepsilon}_{\boldsymbol{\theta}}(\mathbf{v}_T, T).
\label{eq:T-k}
\end{equation}
Then, we apply one step of DDIM inversion from $T-k$ back to $T$:
\begin{equation}
\mathbf{v}'_T = \frac{\boldsymbol{\alpha}_T}{\boldsymbol{\alpha}_{T-k}} \left( \mathbf{v}_{T-k} - \boldsymbol{\sigma}_{T-k} \boldsymbol{\varepsilon}_{\boldsymbol{\theta}}(\mathbf{v}_{T-k}, T-k) \right)
+ \boldsymbol{\sigma}_T \boldsymbol{\varepsilon}_{\boldsymbol{\theta}}(\mathbf{v}_{T-k}, T-k).
\label{eq:v'T}
\end{equation}
Substituting Eq.~\ref{eq:T-k} into Eq.~\ref{eq:v'T} and simplifying, we obtain:
\begin{equation}
\mathbf{v}'_T = \mathbf{v}_T + 
\frac{ \boldsymbol{\alpha}_T \boldsymbol{\sigma}_{T-k} - \boldsymbol{\alpha}_{T-k} \boldsymbol{\sigma}_T }{ \boldsymbol{\alpha}_{T-k} } \cdot
\left[ \boldsymbol{\varepsilon}_{\boldsymbol{\theta}}(\mathbf{v}_T, T) - \boldsymbol{\varepsilon}_{\boldsymbol{\theta}}(\mathbf{v}_{T-k}, T-k) \right].
\label{eq:final}
\end{equation}

The noise prediction network follows the CFG rule:
\begin{equation}
\boldsymbol{\varepsilon}_{\boldsymbol{\theta}}(\mathbf{v}_t, t)
= \left( 1 + \boldsymbol{\omega}_t \right) \cdot \boldsymbol{\varepsilon}_{\boldsymbol{\theta}}(\mathbf{v}_t, t \mid \mathbf{s})
- \boldsymbol{\omega}_t \cdot \boldsymbol{\varepsilon}_{\boldsymbol{\theta}}(\mathbf{v}_t, t \mid \varnothing).
\end{equation}

Assuming $\boldsymbol{\omega}_T = \boldsymbol{\omega}_{\ell}$, $\boldsymbol{\omega}_{T-k} = \boldsymbol{\omega}_{w}$, and invoking first-order Taylor expansion in both time and input, we have:
\begin{align*}
\boldsymbol{\varepsilon}_{\boldsymbol{\theta}}(\mathbf{v}_T, T) &\approx 
\boldsymbol{\varepsilon}_{\boldsymbol{\theta}}\left( \mathbf{v}_{T - \frac{k}{2}}, T - \tfrac{k}{2} \right), \\
\boldsymbol{\varepsilon}_{\boldsymbol{\theta}}(\mathbf{v}_{T-k}, T-k) &\approx 
\boldsymbol{\varepsilon}_{\boldsymbol{\theta}}\left( \mathbf{v}_{T - \frac{k}{2}}, T - \tfrac{k}{2} \right).
\end{align*}

Therefore,
\begin{equation}
\boldsymbol{\varepsilon}_{\boldsymbol{\theta}}(\mathbf{v}_T, T)
- \boldsymbol{\varepsilon}_{\boldsymbol{\theta}}(\mathbf{v}_{T-k}, T-k)
\approx \left( \boldsymbol{\omega}_{\ell} - \boldsymbol{\omega}_{w} \right) \cdot \mathbf{g}_s,
\end{equation}
and inserting into Eq.~\ref{eq:final} yields the closed-form result in Eq.~\ref{eq:simple}.

The expression shows that informative noise $\mathbf{v}'_T$ is obtained by translating the standard Gaussian noise $\mathbf{v}_T$ along the semantic gradient $\mathbf{g}_s$, with a controllable injection magnitude $\boldsymbol{\kappa}$. The gradient direction is derived from the difference in noise predictions under the presence and absence of the structural condition $\mathbf{s}$, and is thus guaranteed to reflect meaningful semantic information in the latent space.

\section{Preference Alignment via Informative Noise: Theoretical Justification}
\label{ap:B}
We analyse the benefit of informative noise from the perspective of user preference alignment in recommendation systems. In particular, we demonstrate that semantically injected informative noise leads to a higher expected preference score compared to standard Gaussian noise.

\vspace{1em}
\begin{theorem}\label{theo:b}
\textit{Let $\mathbf{v}_T \sim \mathcal{N}(\mathbf{0}, \mathbf{I}_d)$ be an initial Gaussian noise vector in the latent space. Let $\mathbf{s} \in \mathbb{R}^{\mathbf{d}_s}$ denote structural semantics, and let $\mathbf{u} \in \mathbb{R}^{\mathbf{d}}$ be a user embedding vector. Let $\mathbf{G} : \mathbb{R}^{\mathbf{d}} \rightarrow \mathbb{R}^{\mathbf{d}}$ be the deterministic generative mapping (e.g., a DDIM sampler over $K$ steps) such that $\mathbf{v}_0 = \mathbf{G}(\mathbf{v}_T)$ yields a generated item embedding.}

\textit{Assume the following conditions:}

\begin{itemize}
  \item $\mathbf{G}$ is $\gamma$-Lipschitz: $\left\| \mathbf{G}(\mathbf{x}) - \mathbf{G}(\mathbf{y}) \right\| \leq \gamma \left\| \mathbf{x} - \mathbf{y} \right\|$ for all $\mathbf{x}, \mathbf{y}$;
  \item The noise predictor $\boldsymbol{\varepsilon}_{\boldsymbol{\theta}}(\cdot, t \mid \cdot)$ is $L$-Lipschitz;
  \item Informative noise is constructed via one re-denoise step:
  \[
  \mathbf{v}_T^{\text{info}} = \mathbf{v}_T + \boldsymbol{\kappa} \cdot \mathbf{g}_s,
  \]
  where the semantic gradient $\mathbf{g}_s \in \mathbb{R}^{\mathbf{d}}$ is defined as:
  \[
  \mathbf{g}_s = \boldsymbol{\varepsilon}_{\boldsymbol{\theta}}\left( \mathbf{v}_{T - \tfrac{k}{2}}, T - \tfrac{k}{2} \mid \mathbf{s} \right)
       - \boldsymbol{\varepsilon}_{\boldsymbol{\theta}}\left( \mathbf{v}_{T - \tfrac{k}{2}}, T - \tfrac{k}{2} \mid \varnothing \right),
  \]
  and $\boldsymbol{\kappa} > 0$ controls injection strength;
  \item There exists $\delta > 0$ such that $\langle \mathbf{u}, \mathbf{g}_s \rangle \ge \delta$, i.e., the semantic direction aligns positively with user preference.
\end{itemize}

\textit{Then the informative-noise-generated embedding $\mathbf{v}_0^{\text{info}} = \mathbf{G}(\mathbf{v}_T^{\text{info}})$ satisfies the following expected preference bound:}
\begin{equation}
\mathbb{E}_{\mathbf{v}_T} \left[ \langle \mathbf{u}, \mathbf{v}_0^{\text{info}} \rangle \right]
\;\ge\;
\mathbb{E}_{\mathbf{v}_T} \left[ \langle \mathbf{u}, \mathbf{v}_0^{\text{std}} \rangle \right]
+ \boldsymbol{\kappa} \delta - \gamma \boldsymbol{\kappa}^2 \left\| \mathbf{u} \right\| \cdot \left\| \mathbf{g}_s \right\|,
\end{equation}
\textit{where $\mathbf{v}_0^{\text{std}} = \mathbf{G}(\mathbf{v}_T)$ is the item embedding generated from standard noise.}
\end{theorem}
\vspace{1em}
\noindent\textit{Proof.}  
We expand the inner product difference:
\begin{equation}
\langle \mathbf{u}, \mathbf{v}_0^{\text{info}} \rangle - \langle \mathbf{u}, \mathbf{v}_0^{\text{std}} \rangle
= \left\langle \mathbf{u}, \mathbf{G}(\mathbf{v}_T + \boldsymbol{\kappa} \mathbf{g}_s) - \mathbf{G}(\mathbf{v}_T) \right\rangle.
\end{equation}

Using first-order Taylor expansion of $\mathbf{G}$ around $\mathbf{v}_T$, we have:
\begin{equation}
\mathbf{G}(\mathbf{v}_T + \boldsymbol{\kappa} \mathbf{g}_s)
\approx \mathbf{G}(\mathbf{v}_T) + \boldsymbol{\kappa} \mathbf{J}_{\mathbf{G}}(\mathbf{v}_T) \mathbf{g}_s + \mathbf{R},
\end{equation}
where $\mathbf{J}_{\mathbf{G}}(\mathbf{v}_T)$ is the Jacobian of $\mathbf{G}$ and $\mathbf{R}$ is the residual. By Lipschitz continuity and bounded Hessian curvature, the remainder satisfies:
\begin{equation}
\left\| \mathbf{R} \right\| \le \frac{1}{2} \gamma \lambda_{\max}(\mathbf{H}_{\mathbf{G}}) \cdot \boldsymbol{\kappa}^2 \left\| \mathbf{g}_s \right\|^2.
\end{equation}

Hence the inner product becomes:
\begin{equation}
\langle \mathbf{u}, \mathbf{v}_0^{\text{info}} \rangle - \langle \mathbf{u}, \mathbf{v}_0^{\text{std}} \rangle
\ge \boldsymbol{\kappa} \left\langle \mathbf{J}_{\mathbf{G}}^\top(\mathbf{v}_T) \mathbf{u}, \mathbf{g}_s \right\rangle
- \left\| \mathbf{u} \right\| \cdot \left\| \mathbf{R} \right\|.
\end{equation}

If $\mathbf{G}$ is approximately linear near $\mathbf{v}_T$, as in many encoder-decoder diffusion frameworks, we approximate $\mathbf{J}_{\mathbf{G}}^\top(\mathbf{v}_T) \mathbf{u} \approx \mathbf{u}$. Then,
\begin{equation}
\langle \mathbf{u}, \mathbf{v}_0^{\text{info}} \rangle - \langle \mathbf{u}, \mathbf{v}_0^{\text{std}} \rangle
\ge \boldsymbol{\kappa} \langle \mathbf{u}, \mathbf{g}_s \rangle - \gamma \boldsymbol{\kappa}^2 \left\| \mathbf{u} \right\| \cdot \left\| \mathbf{g}_s \right\|.
\end{equation}

Taking expectation over $\mathbf{v}_T$, and using the assumption $\langle \mathbf{u}, \mathbf{g}_s \rangle \ge \delta$, we obtain the claimed bound.

\vspace{1em}
\noindent\textbf{Discussion.}  
The result shows that informative noise $\mathbf{v}_T^{\text{info}}$, as constructed through re-denoising, produces item embeddings with strictly better expected preference score compared to those generated from standard noise, provided the semantic direction $\mathbf{g}_s$ aligns positively with user intent. Moreover, by choosing
\begin{equation}
\boldsymbol{\kappa}^* = \frac{\delta}{2 \gamma \left\| \mathbf{u} \right\| \cdot \left\| \mathbf{g}_s \right\|},
\end{equation}
the improvement in expected inner product is maximized at:
\begin{equation}
\Delta^* = \frac{\delta^2}{4 \gamma \left\| \mathbf{u} \right\| \cdot \left\| \mathbf{g}_s \right\|}.
\end{equation}

This offers theoretical guidance for selecting optimal injection strength $\boldsymbol{\kappa}$, e.g., via adjusting $\boldsymbol{\omega}_{\ell} - \boldsymbol{\omega}_{w}$ or the DDIM step size $k$. It confirms that informative noise is not only a semantically meaningful modification but also improves preference alignment in a mathematically controllable and verifiable way.
\begin{figure}[!t]
    \centering
    \captionsetup{skip=0pt}
    \includegraphics[width=\linewidth]{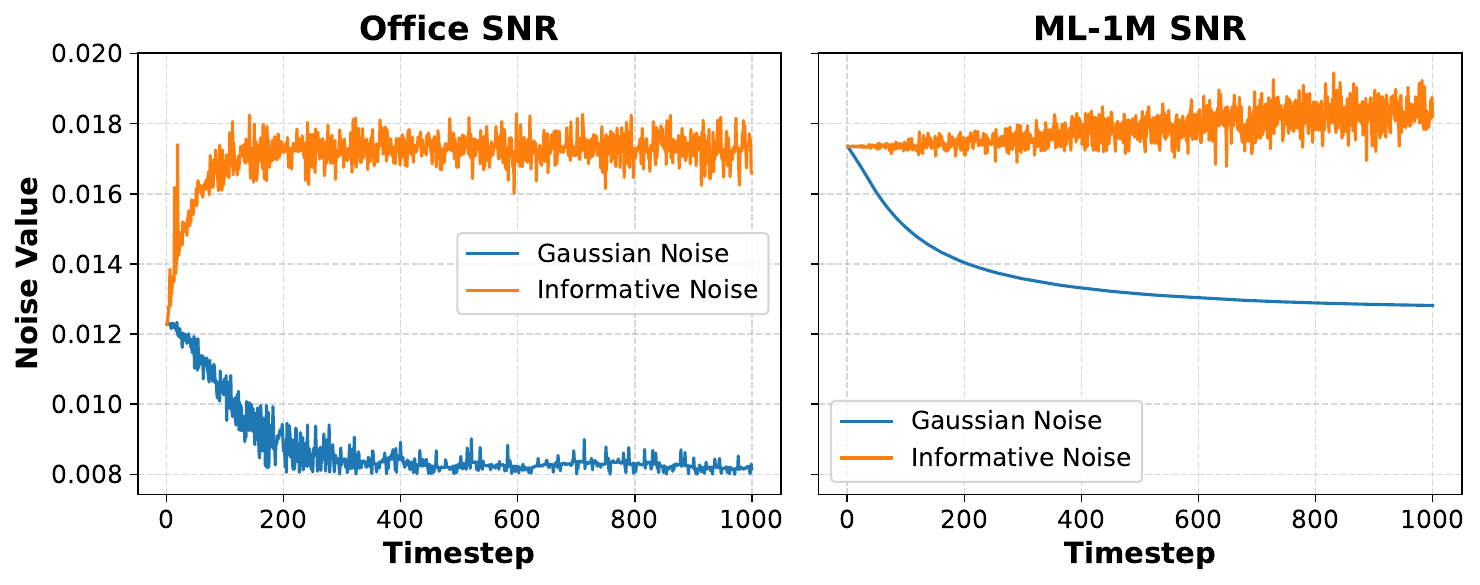}
    \caption{SNR Comparison on Amazon-Office and ML-1M}
    \label{fig:snr_ap}
\end{figure}
\section{Hyperparameter Analysis}

\begin{figure*}[!t]
    \centering
    \captionsetup{skip=0pt}
    \includegraphics[width=\textwidth]{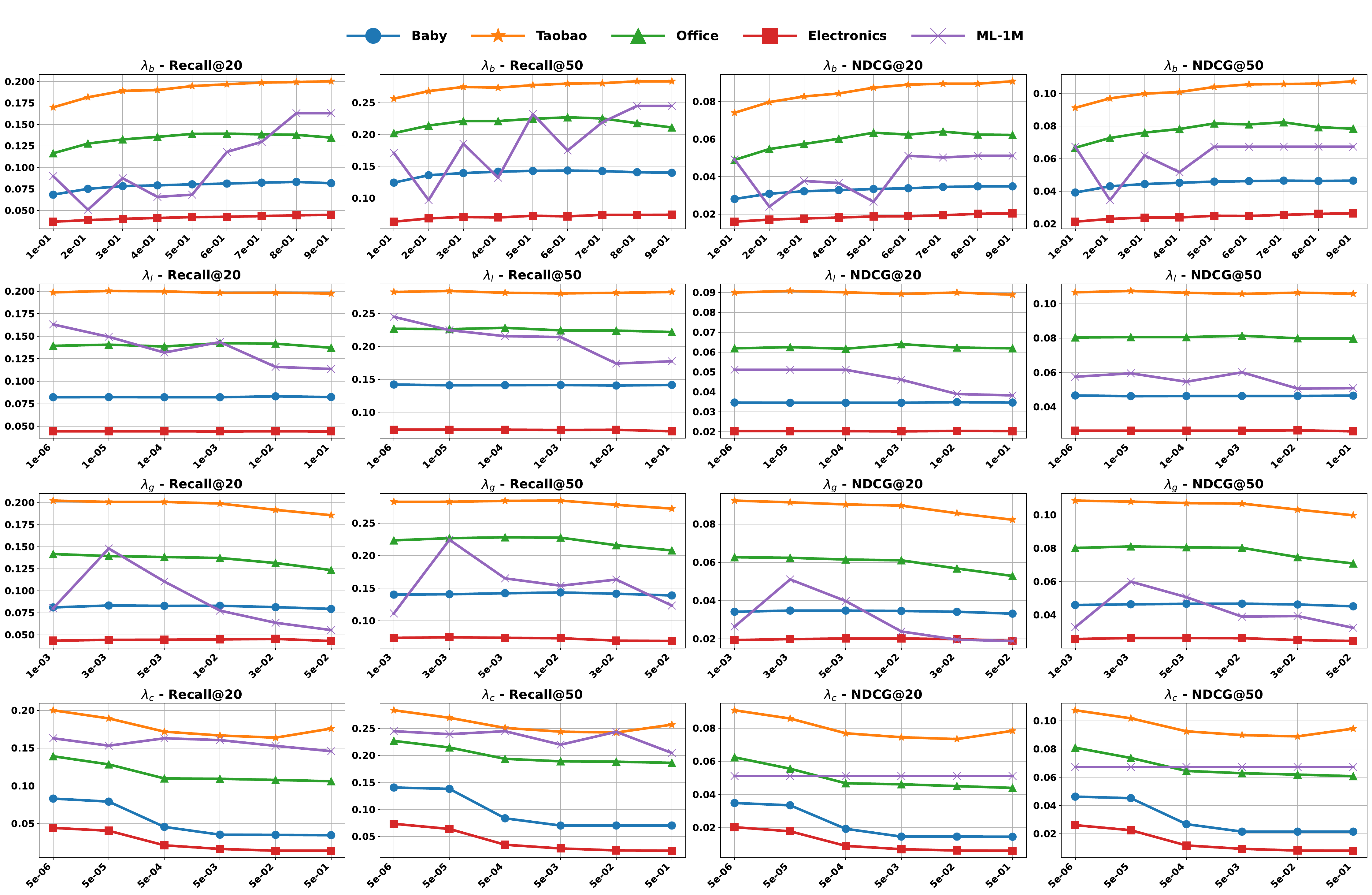}
    \caption{Hyperparameter analysis across five datasets}
    \label{fig:hy}
\end{figure*}

To investigate the impact of key hyperparameters, we conduct experiments on five benchmark datasets: Baby, Office, Taobao, Electronics and ML-1M. The studied hyperparameters include: (1) the BPR loss coefficient $\lambda_{b}$; (2) the contrastive learning coefficient $\lambda_{con}$; (3) the coefficient of collaboration balance loss $\lambda_{l}$; and (4) the regularization loss coefficient $\lambda_{g}$. The results are shown in Figure~\ref{fig:hy}.

We observe that increasing the BPR loss coefficient $\lambda_b$ generally improves performance across all datasets, especially for Taobao and Office. This suggests that a moderate to high value of $\lambda_b$ contributes positively to learning effective user-item representations. Regarding the collaboration balance coefficient $\lambda_l$, its influence appears dataset-dependent. ML-1M and Office datasets show the best performance at mid-range values (around $1e$-$3$ or $1e$-$4$), indicating the need to carefully balance the generation and preference learning for different data sparsity levels.

For the regularization coefficient $\lambda_{g}$, we find that lower values (e.g., $3e$-$3$ to $5e$-$3$) yield better performance, especially on dense datasets like ML-1M. Over-regularization tends to suppress learning effect and lead to degraded performance. Finally, for the contrastive learning coefficient $\lambda_{con}$, smaller values (such as $5e$-$3$) consistently lead to improved metrics across datasets, suggesting that while contrastive signals are beneficial, overly strong contrastive losses might dominate the learning and hurt generalization, particularly on sparse datasets such as Baby and Electronics.

\section{SNR Comparison Between Gaussian Noise and Informative Noise}
We further visualize the SNR comparison curves for the Amazon-Office and ML-1M datasets in Figure \ref{fig:snr_ap}, which provides additional evidence that our InfoDCL is capable of effectively capturing semantic information across diverse datasets.
\section{Related Work}
\subsection{Diffusion Based Recommendation}
Diffusion models\cite{CFDG,GenerativeModeling} have gained substantial attention since DDPM \cite{ddpm}, which laid the groundwork for learning data distributions via iterative addition of Gaussian noise followed by progressive denoising. Building on this foundation, subsequent research has sought to improve sampling efficiency and flexibility. Notably, non-Markovian formulations \cite{ddim} achieved faster generation by modifying the diffusion trajectories, while conditional generation approaches \cite{diffbeatgans} enhanced output controllability using external signals such as class labels or textual descriptions. These advancements have enabled diffusion-based techniques to extend into domains like recommendation systems.

In the context of recommender systems, DiffRec \cite{diffrec} was among the first to treat user-item interaction modeling as a denoising task. DreamRec \cite{dreamrec} further advanced this line of work by incorporating sequential behavior, introducing time-aware weights to better capture dynamic user preferences through recent interactions. More recent developments have explored refined conditioning strategies and physically inspired methodologies. For example, DDRM \cite{ddrm} implements mutual conditioning between users and items, allowing both sides to co-evolve during the reverse diffusion process through joint gradient updates. In parallel, GiffCF \cite{giffcf} proposes a diffusion mechanism grounded in graph signal processing, simulating a heat diffusion process on the interaction graph via the Laplacian to propagate affinity signals. In conclusion, these contributions underscore the promise of diffusion models in capturing intricate patterns within user-item relationships.

\subsection{Contrastive Learning in Collaborative Filtering}
In recent years, contrastive learning has emerged as a powerful paradigm in representation learning, leveraging instance-level discrimination to extract robust embeddings. In collaborative filtering, early methods like SGL~\cite{sgl} used self-supervised node-level contrastive objectives over user--item graphs, enhancing representation robustness via graph augmentations such as node/edge dropout and random walk. NCL~\cite{ncl} extended this idea by leveraging multi-hop structural neighbors and semantic prototypes as contrasting views, enriching representation learning through both topology and semantics. Following this trajectory, SimGCL~\cite{simgcl} showed that perturbing embeddings directly (rather than graph structure augmentation) could achieve competitively better performance in recommendation, promoting both geometric uniformity and training efficiency.

SCCF~\cite{sccf} further unified graph convolution and contrastive learning under a theoretical framework, proposing a simple GCN-free design that achieves high-order connectivity modeling via contrastive objectives alone. Complementarily, DimCL~\cite{dimcl} introduced dimension-aware augmentation by analyzing the learning dynamics of latent dimensions, enabling contrastive learning to focus on harder-to-learn features for improved representation robustness. These advances highlight the adaptability of contrastive learning in modeling intricate user--item relationships and its potential to address sparsity and generalization challenges in collaborative filtering. By effectively regularizing the latent space to distinguish hard negatives, these paradigms ensure that the learned representations remain discriminative even in highly sparse scenarios.

\end{document}